\documentclass[aps,PRB,reprint,superscriptaddress,preprintnumbers]{revtex4-2}
\usepackage[T1]{fontenc}
\usepackage{times}
\usepackage{graphicx,color}
\usepackage{amsfonts,amsmath,amssymb,amsbsy}
\usepackage[colorlinks=true, linkcolor=blue, citecolor=blue, urlcolor=blue]{hyperref}
\usepackage{float}

\begin{document}

\title{Universal quantum computer based on Carbon Nanotube Rotators}
\author{Motohiko Ezawa}
\affiliation{Department of Applied Physics, The University of Tokyo, 7-3-1 Hongo, Tokyo
113-8656, Japan}
\author{Shun Yasunaga}
\affiliation{Department of Electrical Engineering, University of Tokyo, Hongo 7-3-1,
113-8656, Japan}
\author{Tetsuya Iizuka}
\affiliation{Department of Electrical Engineering, University of Tokyo, Hongo 7-3-1,
113-8656, Japan}
\author{Akio Higo}
\affiliation{Department of Electrical Engineering, University of Tokyo, Hongo 7-3-1,
113-8656, Japan}
\author{Yoshio Mita}
\affiliation{Department of Electrical Engineering, University of Tokyo, Hongo 7-3-1,
113-8656, Japan}

\begin{abstract}
We propose a universal quantum computer based on a chain of carbon nanotube
rotators where one metallic plate is attached to each rotator. The dynamical
variable is the rotational angle $\phi$. The attached plate connected to
ground electrostatically interacts with two fixed plates. Two angle
positions $\phi =0,\pi$ are made stable by applying a voltage difference
between the attached plate and the two fixed plates. We assign $\phi =0$ and 
$\pi $ to the qubit states $|0\rangle$ and $|1\rangle $. Then, considering a
chain of rotators, we construct the arbitrary phase-shift gate, the NOT gate
and the Ising gate, which constitute a set of universal quantum gates. They
are executed by controlling the voltage between various plates.
\end{abstract}

\date{\today }
\maketitle

\section{Introduction}

The Moor law is a fundamental roadmap of integrated circuits, which dictates
that the number of elements increases exponentially as a function of year.
It also means that the size of an element must become exponentially small.
However, there is an intrinsic limit of an element, which is the size of
atoms of the order of 1nm. It gives a limit to the Moor law. The quantum
computer is expected to be a solution to overcome it\cite%
{Feynman,DiVi,Nielsen}. Quantum computers are based on qubits, where the
superposition of the quantum states $|0\rangle $ and $|1\rangle $ is used.
Various methods have been proposed such as superconductors\cite{Nakamura},
photonic systems\cite{Knill}, ion trap\cite{Cirac}, nuclear magnetic
resonance\cite{Vander,Kane}, quantum dots\cite{Loss}, skyrmions\cite%
{Psa,SkBit} and merons\cite{MeronBit}. Nanomechanical systems are also
applicable to quantum computers\cite{Rips,Rips2,Pisto,NEMS}.

Quantum algorithm is decomposed into a sequential application of quantum
gates. The Solovay-Kitaev theorem assures that only three quantum gates, the 
$\pi /4$ phase-shift gate, the Hadamard gates and the CNOT gate, are enough
for universal quantum computations\cite{Deutsch,Dawson,Universal}.
Alternatively, the arbitrary phase-shift gate, the NOT gate and the Ising
gate constitute a set of universal quantum gates as well.

Nano-electromechanical systems (NEMS)\cite{Cra,Eki} have various industrial
applications. A nanorotator based on a carbon nanotube has been
experimentally realized\cite{Fenn,Barre,CaiRotor,CaiCarbon}. Especially, a
double-wall nanotube structure acts as a nanomotor\cite%
{Kolmo,Barre,Bour,Xu,CaiRotor,CaiNano,CaiCarbon,Zamb}. A carbon nanotube can
be metallic depending on the chirality of a nanotube\cite{Saito}. In
addition, it is possible to attach a metallic plate to a nanotube\cite%
{Fenn,Bour}. Quantum effects are experimentally observed in NEMS\cite%
{Blenco,Poot,Slowik}.

\begin{figure}[t]
\centerline{\includegraphics[width=0.48\textwidth]{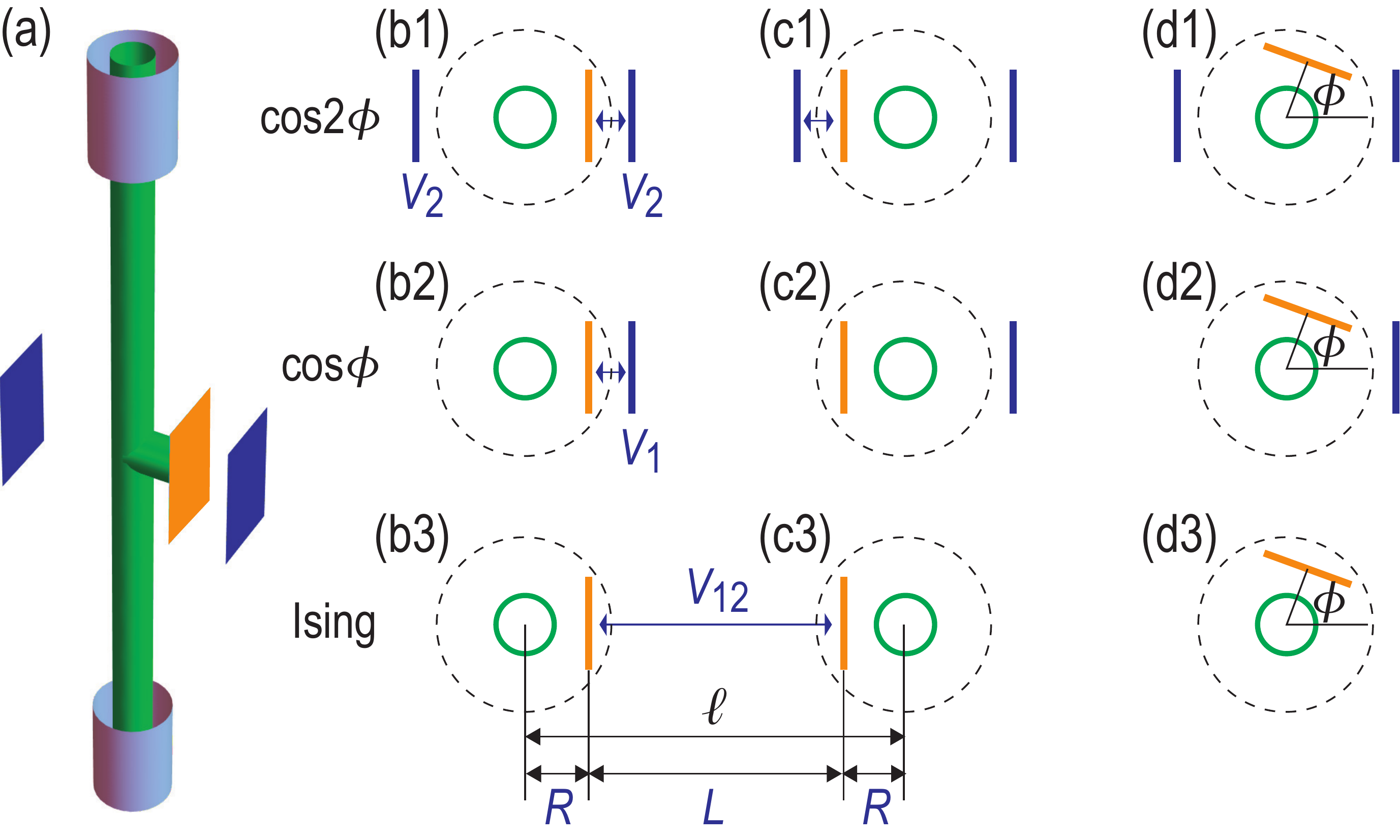}}
\caption{(a) Illustration of a nanotube rotator suspended by the double-wall
nanotube structures at the top and the bottom. One metallic plate (in
orange) is attached to the nanotube. Two metallic plates (in blue) are fixed
to the outer system. We connect the inner plate to ground. When we give a
voltage to the two outer plates, the $\cos 2\protect\phi $ potential is
induced. When we give a voltage to one of the two outer plates, the $\cos 
\protect\phi $ potential is induced. On the other hand, when we give a
voltage between the two plates attached to two nanotubes, the Ising
interaction is induced. (b) The configuration of a rotator with $\protect%
\phi =0$, representing the qubit state $\left\vert 0\right\rangle $. (c) The
configuration of a rotator with $\protect\phi =\protect\pi $, representing
the qubit state $\left\vert 1\right\rangle $. (d) The configuration of a
rotator with a generic angle $\protect\phi $. Dotted circles denote the
rotator parts. }
\label{FigIllust}
\end{figure}

In this paper, we propose a universal quantum computer by constructing a set
of universal quantum gates. We prepare a rotator based on a double-wall
nanotube as illustrated in Fig.\ref{FigIllust}(a). We attach one metallic
plate to the inner nanotube. It is possible to materialize such a
nanorotator by using the present techniques\cite{Fenn,Bour}. Then, we align
these rotators along a line with equal spacing, which is the main
configuration of our proposal. We explicitly design the arbitrary
phase-shift gate, the NOT gate and the Ising gate. They are controlled by
the voltage between two plates.

\section{Model}

\subsection{Carbon nanotube rotator}

We consider a rotator whose dynamical variable is the rotational angle $\phi 
$ with the potential energy given by\cite{Bour} 
\begin{equation}
W_{2}(\phi )=-A\cos 2\phi .  \label{W2}
\end{equation}%
There are two stable angles $\phi =0$ and $\pi $, which we regard to form
one-qubit states \{$|0\rangle ,|1\rangle $\}. In addition, we introduce a
potential term given by\cite{Bour}%
\begin{equation}
W_{1}(\phi )=-B\cos \phi ,  \label{W1}
\end{equation}%
which resolves the degeneracy between the states $|0\rangle $ and $|1\rangle 
$.

The Schr\"{o}dinger equation for one rotator reads%
\begin{equation}
i\hbar \frac{d}{dt}\psi \left( t\right) =H\psi \left( t\right) ,
\label{Schroe}
\end{equation}%
with the Hamiltonian given by%
\begin{equation}
H=-\frac{\hbar ^{2}}{2\mu r^{2}}\frac{\partial }{\partial \phi ^{2}}-A\cos
2\phi -B\cos \phi ,  \label{HamilA}
\end{equation}%
where $\mu $ is the inertia of the rotator and $r$ is the radius of the
rotator. The eigenequation reads%
\begin{equation}
H\psi =E\psi .  \label{HEPsi}
\end{equation}

The Hamiltonian (\ref{HamilA}) may be materialized by a rotator, which is
made of a nanotube (in green) supported by the double-wall nanotube
structures at the top and the bottom, as illustrated in Fig.\ref{FigIllust}%
(a). We attach one metal plate (in orange) to the nanotube, which we call
the inner plate. Then, we introduce two metal plates (in blue) fixed to an
outer system, which we call the outer plates.

1) We connect the inner plate to ground. It materializes the potential
energy (\ref{W2}) when we give a voltage $\varpropto V_{2}$ to the two outer
plates. There are two stable angles $\phi =0$ and $\pi $, as illustrated in
Fig.\ref{FigIllust}(b1) and (c1). We also illustrate a rotor with a generic
angle $\phi $ in Fig.\ref{FigIllust}(d1).

2) We connect the inner plate to ground. It materializes the potential
energy (\ref{W1}) when we give a voltage $\varpropto V_{1}$ to one of the
two outer plates. There is one stable angle $\phi =0$, as illustrated in Fig.%
\ref{FigIllust}(b2). We also illustrate a rotor with a generic angle $\phi $
in Fig.\ref{FigIllust}(d2).

\subsection{Whittaker--Hill Equation}

The Schr\"{o}dinger equation (\ref{Schroe}) may be rewritten in a
dimensionless form as%
\begin{equation}
i\frac{d}{d\mathcal{\tau }}\psi \left( \phi ,\tau \right) =\mathcal{H}\psi
\left( \phi ,\tau \right) ,
\end{equation}%
with the dimensionless Hamiltonian,%
\begin{equation}
\mathcal{H=}-\frac{d^{2}}{d\phi ^{2}}+\mathcal{V(}\phi ),  \label{HamilB}
\end{equation}%
and the dimensionless potential,%
\begin{equation}
\mathcal{V\mathcal{(}\phi )=}-\mathcal{V}_{2}\cos 2\phi -\mathcal{V}_{1}\cos
\phi ,  \label{PotenB}
\end{equation}%
where%
\begin{align}
\mathcal{\tau }& =\frac{\hbar }{2\mu r^{2}}t,\quad \varepsilon =\frac{2\mu
r^{2}}{\hbar ^{2}}E,\quad   \notag \\
\mathcal{V}_{2}& =\frac{2\mu r^{2}}{\hbar ^{2}}A,\quad \mathcal{V}_{1}=\frac{%
2\mu r^{2}}{\hbar ^{2}}B.
\end{align}%
The dimensionless quantity $\mathcal{V}_{2}$ ($\mathcal{V}_{1}$) is given in
terms of the voltage difference $V_{1}$($V_{2}$) between the inner plate and
the two (one) outer plates,%
\begin{equation}
\mathcal{V}_{i}=\frac{1}{2}CV_{i}^{2},
\end{equation}%
where $C$ is the capacitance of the inner-outer plate system, as indicated
by $\cos 2\phi $ ($\cos \phi $) in Fig.\ref{FigIllust}(a). We show the
potential $\mathcal{V}(\phi )$\ for $\mathcal{V}_{1}=0$\ in Fig.\ref%
{FigMathieuWave}(a1)\ and for $\mathcal{V}_{1}=1$\ in Fig.\ref%
{FigMathieuWave}(b1) by setting $\mathcal{V}_{2}=20$.

\begin{figure}[t]
\centerline{\includegraphics[width=0.48\textwidth]{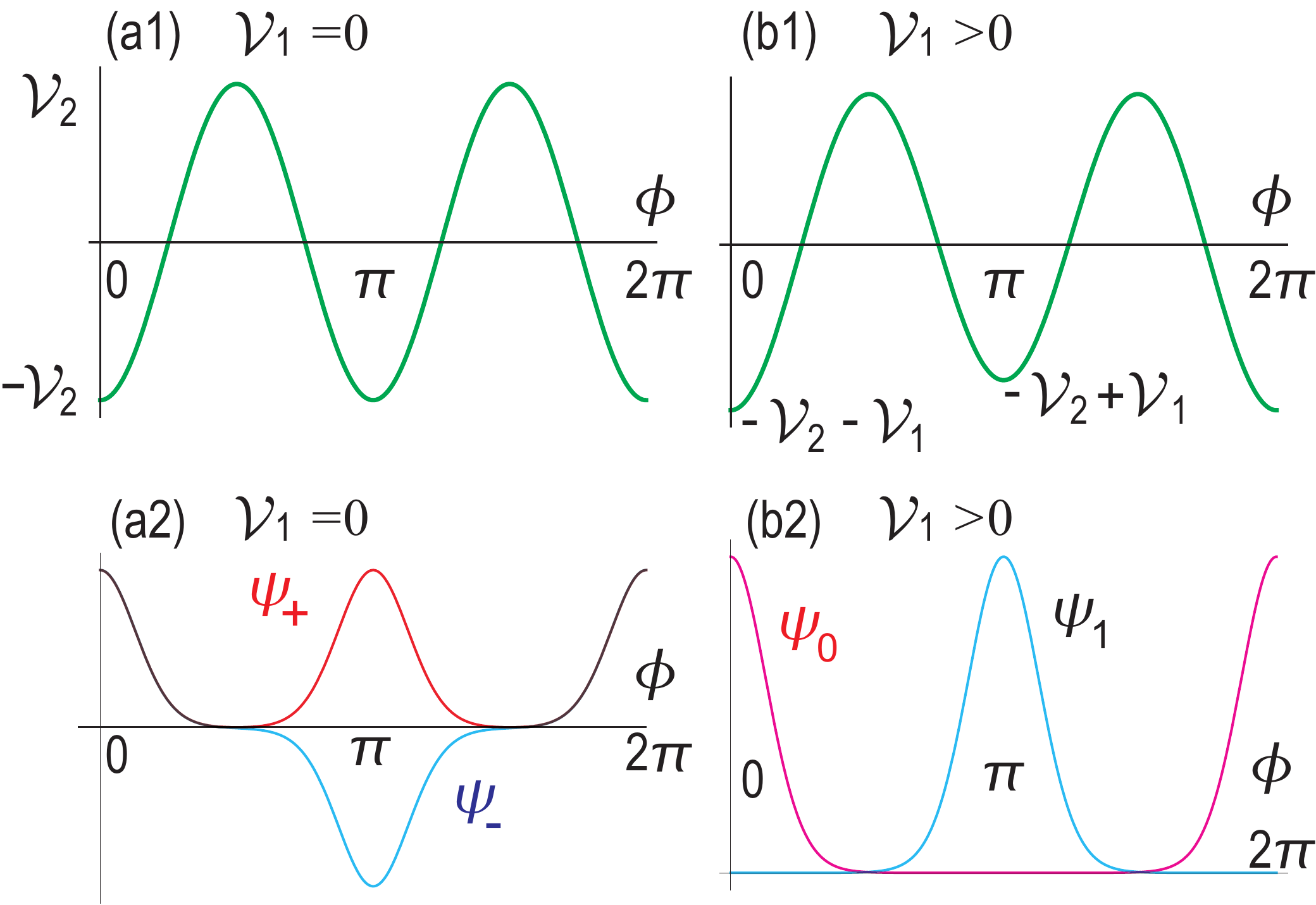}}
\caption{(a1) Potential energy as a function of $\protect\phi $, where there
are two minima at $\protect\phi =0$ and $\protect\pi $; (a2) wave functions
as a function of $\protect\phi $, where the magenta curve indicates the
symmetric ground state $\protect\psi _{+}$ and the cyan curve indicates the
antisymmetric first-excited state $\protect\psi _{-}$. Here, we have set $%
\mathcal{V}_{1}=0$ and $\mathcal{V}_{2}=20$. (a2) Potential energy as a
function of $\protect\phi $, where there is only one minimum at $\protect%
\phi =0$; (b2) wave functions as a function of $\protect\phi $, where the
magenta curve indicates the eigenfunction $\protect\psi _{0}$ of the state $%
|0\rangle $ and the cyan curve indicates $\protect\psi _{1}$ of $|1\rangle $%
. Here, we have set $\mathcal{V}_{1}=1$ and $\mathcal{V}_{2}=20$.}
\label{FigMathieuWave}
\end{figure}

The eigenequation $\mathcal{H}\psi =\varepsilon \psi $ reads%
\begin{equation}
\left[ \frac{d^{2}}{d\phi ^{2}}+\mathcal{V}_{2}\cos 2\phi +\mathcal{V}%
_{1}\cos \phi \right] \psi =-\varepsilon \psi .  \label{EqWH}
\end{equation}%
This is the Whittaker--Hill equation, which is reduced to the Mathieu
equation for $\mathcal{V}_{1}=0$.

\subsection{Strong potential limit}

As the basic picture of the present model, we require the dominant role of
the cosine potential $\cos 2\phi $ to generate two-fold degenerated ground
states at $\phi =0$ and $\pi $, which we regard to form one-qubit states \{$%
|0\rangle ,|1\rangle $\}. On the other hand, we use the cosine potential $%
\cos \phi $ to make gate operations. Hence, we consider the regime where $%
\mathcal{V}_{2}\gg \mathcal{V}_{1}\geq 0$. It is possible to derive the
analytical solutions around these two points for sufficiently large $%
\mathcal{V}_{2}$. The Whittaker-Hill equation (\ref{EqWH}) is approximated as%
\begin{align}
& \left[ \frac{d^{2}}{d\phi ^{2}}+\left( \mathcal{V}_{2}+\left( -1\right)
^{q}\mathcal{V}_{1}\right) -\left( 2\mathcal{V}_{2}+\left( -1\right) ^{q}%
\frac{\mathcal{V}_{1}}{2}\right) \phi _{q}^{2}\right] \psi _{q}  \notag \\
& \hspace{4cm}=\varepsilon _{q}\psi _{q},  \label{ApproWH}
\end{align}%
where $\phi _{q}=\phi -q\pi $ with $q=0,1$.

By solving Eq.(\ref{ApproWH}), the wave function $\psi _{q}$ is given by%
\begin{equation}
\psi _{q}\left( \phi _{q}\right) =\left( \frac{2a}{\pi }\right)
^{1/4}e^{-\alpha _{q}\phi _{q}^{2}},  \label{Psi0}
\end{equation}%
with 
\begin{equation}
\alpha _{q}=\sqrt{\frac{\mathcal{V}_{2}}{2}+\left( -1\right) ^{q}\frac{%
\mathcal{V}_{1}}{8}}.
\end{equation}%
The energy of the state (\ref{Psi0}) is%
\begin{equation}
\varepsilon _{q}=-\left( \mathcal{V}_{2}+\left( -1\right) ^{q}\mathcal{V}%
_{1}\right) +\sqrt{2\mathcal{V}_{2}+\left( -1\right) ^{q}\frac{\mathcal{V}%
_{1}}{2}}.
\end{equation}%
When $\mathcal{V}_{1}>0$, the ground state is given by $\psi _{0}$, and the
first excited state by $\psi _{1}$. The wave function $\psi _{q}$\ describes
the one-qubit state $|q\rangle $\ with the qubit variable $q=0,1$.

When $\mathcal{V}_{1}=0$, these two states are degenerate. However, an
energy splitting occurs due to the difference between the Whittaker-Hill
equation (\ref{EqWH}) and the approximated equation (\ref{ApproWH}). Then,
due to the mixing, the ground state and the first-excited state wave
functions turn out to be the symmetric function $\psi _{+}$ and the
antisymmetric function $\psi _{-}$,%
\begin{equation}
\psi _{+}=\frac{\psi _{0}+\psi _{1}}{\sqrt{2}},\quad |\psi _{-}\rangle
\equiv \frac{\psi _{0}-\psi _{1}}{\sqrt{2}},
\end{equation}%
however small the energy splitting is.

\subsection{Numerical analysis}

The Whittaker--Hill equation is solved by making a Fourier series expansion,%
\begin{equation}
\psi =\sum_{n=-\infty }^{\infty }\alpha _{n}e^{in\phi }.  \label{Fourie}
\end{equation}%
The coefficient $\alpha _{n}$ is determined by solving a set of
eigenequations,%
\begin{equation}
-n^{2}\alpha _{n}+\frac{\mathcal{V}_{2}}{2}\left( \alpha _{n+2}+\alpha
_{n-2}\right) +\frac{\mathcal{V}_{1}}{2}\left( \alpha _{n+1}+\alpha
_{n-1}\right) =\varepsilon \alpha _{n}.
\end{equation}%
This is summarized in the matrix form, 
\begin{equation}
\sum_{m}M_{nm}\alpha _{m}=\varepsilon \alpha _{n}.
\end{equation}%
We have numerically solved this matrix equation by introducing a cut off as
in%
\begin{equation}
\psi =\sum_{n=-N}^{N}\alpha _{n}e^{in\phi },  \label{Alpha}
\end{equation}%
with a certain integer $N$. We have checked that it is enough to take $N=8$.
Here we use $N=12$.

\begin{figure}[t]
\centerline{\includegraphics[width=0.48\textwidth]{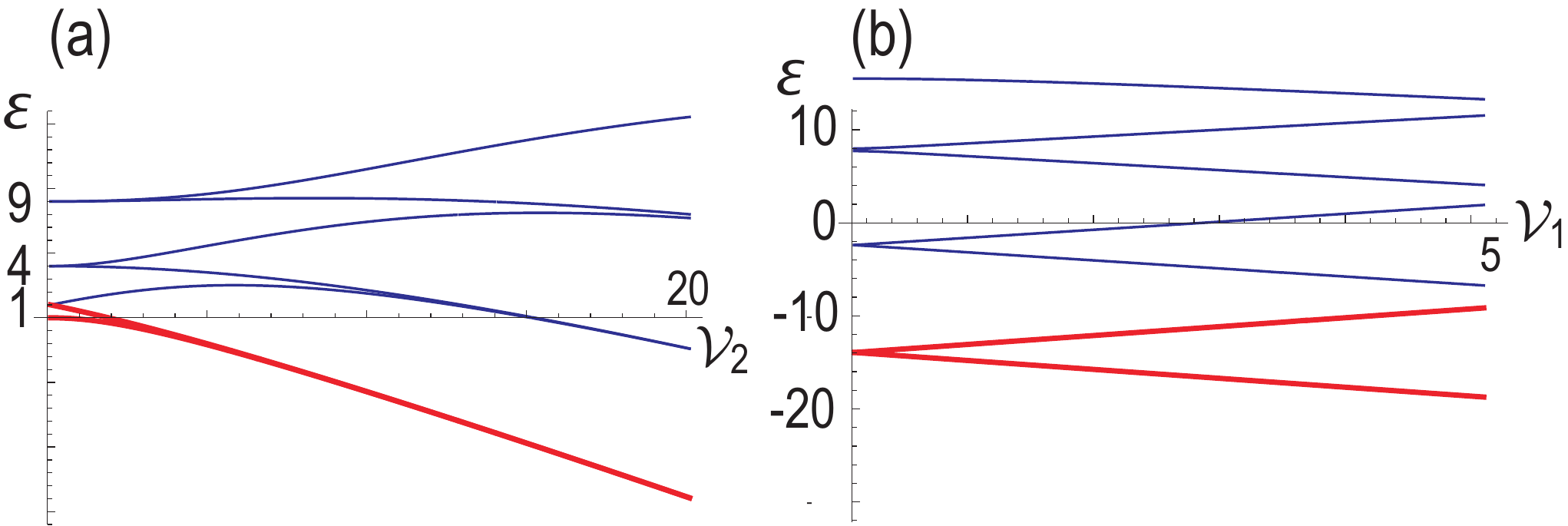}}
\caption{ (a) Eigenenergy $\protect\varepsilon $ as a function of $\mathcal{V%
}_{2}$ by setting $\mathcal{V}_{1}=0$. (b) Eigenenergy $\protect\varepsilon $
as a function of $\mathcal{V}_{1}$ by setting $\mathcal{V}_{2}=20$. The
lowest two energy levels are colored in red, while the other energy levels
are colored in blue.}
\label{FigMathieuEnergy}
\end{figure}

We show the energy spectrum as a function of $\mathcal{V}_{2}$ by setting $%
\mathcal{V}_{1}=0$ in Fig.\ref{FigMathieuEnergy}(a). We are concerned about
the lowest two energy levels indicated in red, which are well separated from
all the other. The energy is two-fold degenerated in the limit of $\mathcal{V%
}_{2}\rightarrow \infty $. Their wave functions are given by the symmetric
function $\psi _{+}$ and the antisymmetric function $\psi _{-}$, as shown in
Fig.\ref{FigMathieuWave}(a2).

We show the energy spectrum as a function of $\mathcal{V}_{1}$ by setting $%
\mathcal{V}_{2}=20$ in Fig.\ref{FigMathieuEnergy}(b). The almost two-fold
degenerated energy levels split linearly as a function of $\mathcal{V}_{1}$.
The ground state and the first-excited state wave functions are $\psi _{0}$
and $\psi _{1}$, which are localized at $\phi =0$ and $\pi $, as shown in
Fig.\ref{FigMathieuWave}(b2) for the case of $\mathcal{V}_{2}=20$ and $%
\mathcal{V}_{1}=1$.

\subsection{Quantum tunneling}

We use the two states $\left\vert 0\right\rangle $ and $\left\vert
1\right\rangle $ as the one-qubit states. Because these two states are
degenerate when $\mathcal{V}_{1}=0$, one may wonder if they are naturally
mixed by quantum tunneling. Then, the life time of a qubit state is too
short. However, this is not the case. During gate operations we keep $%
\mathcal{V}_{2}$ quite large to keeps the cosine potential well defined, as
generates a quite large barrier between these two states.

Quantum tunneling is estimated by means of the WKB approximation. The
tunneling rate $\Gamma $ is given by%
\begin{equation}
\Gamma =e^{-2\gamma },\quad \gamma \equiv \frac{1}{\hbar }\int_{0}^{\pi
}d\phi \left\vert \sqrt{2\mu r^{2}\left( \mathcal{V}_{2}+\mathcal{V}%
_{2}\left( \phi \right) \right) }\right\vert .  \label{Tunnel}
\end{equation}%
By setting $\mathcal{V}_{2}\left( \phi \right) =-\mathcal{V}_{2}\cos 2\phi $%
, we find%
\begin{equation}
\gamma =\frac{4\sqrt{\mu r^{2}\mathcal{V}_{2}}}{\hbar }.
\end{equation}%
Hence, the quantum tunneling is exponentially small as a function of the
applied voltage $V_{2}$. It is estimated that $\gamma =10^{6}\sim 10^{8}$
for $V_{2}=$1mV$\sim $100mV, where we have used that the inertia $\mu r^{2}$
is 10$^{-30}$kgm$^{2}$\cite{Yuz}. See Sec.\ref{SecDiscuss} with respect to
these parameters.

\section{Qubit operations}

\subsection{Initialization}

The present system is composed of a chain of nanotube rotators, each of
which is subject to the cosine potential $\cos 2\phi _{n}$ as in Fig.\ref%
{FigMathieuWave}(a1) by requiring $\mathcal{V}_{1}=0$. The ground states are 
$N$-qubit states $\left\vert q_{1}q_{2}\cdots q_{N}\right\rangle $ with $%
q_{n}=0,1$ for $n=1,2,\cdots ,N$. The initialization to the state $%
\left\vert 00\cdots 0\right\rangle $ is necessary for quantum computations.
It is done by the annealing method. First, we start with a high temperature,
where the rotational angle is random. Then, we cool down the sample slowly
by applying a voltage $\mathcal{V}_{1}$ to all the outer plates in the
right-hand side of the rotators. The rotators tend to the angle $\phi _{n}=0$
in order to minimize the electrostatic energy. As a result, all of the
rotators have the angle $\phi _{n}=0$, which corresponds to the state $%
\left\vert 00\cdots 0\right\rangle $.

\subsection{One-qubit gates}

We construct quantum gates for universal quantum computations. It is enough
to design the arbitrary phase-shift gate and the NOT gate with respect to
one-qubit gates. These gates are realized by varying the parameters $%
\mathcal{V}_{1}$ and $\mathcal{V}_{2}$ in the potential (\ref{PotenB}). We
investigate the quantum dynamics governed by%
\begin{equation}
i\frac{d}{d\tau }\psi \left( \phi ,\tau \right) =\mathcal{H}\left( \tau
\right) \psi \left( \phi ,\tau \right) ,
\end{equation}%
or equivalently,%
\begin{equation}
i\frac{d}{d\tau }\alpha _{n}\left( \tau \right) =\sum_{m}M_{nm}\left( \tau
\right) \alpha _{m}\left( \tau \right) ,
\end{equation}%
in terms of the coefficients $\alpha _{n}$ in Eq.(\ref{Fourie}). We
numerically solve these differential equations in the following.

In the present instance, we assume that either $\mathcal{V}_{1}$ or $%
\mathcal{V}_{2}$ is time dependent during a gate operation. We consider a
quantum gate operation satisfying%
\begin{equation}
\mathcal{H}(\tau _{\text{final}})=\mathcal{H}(\tau _{\text{initial}}).
\end{equation}%
Namely, we tune so that $\mathcal{V}_{2}(\tau _{\text{final}})=\mathcal{V}%
_{2}(\tau _{\text{initial}})$ and $\mathcal{V}_{1}(\tau _{\text{final}})=%
\mathcal{V}_{1}(\tau _{\text{initial}})$. Then, the wave function after the
gate operation is expanded by the superposition of the two gaussian
functions (\ref{Psi0}), which enables us to determine the coefficients of $%
\left\vert 0\right\rangle $ and $\left\vert 1\right\rangle $.

This process is represented by a unitary matrix $U$ from the initial state $%
(\left\vert 0\right\rangle _{\text{initial}},\left\vert 1\right\rangle _{%
\text{initial}})$ to the final state $(\left\vert 0\right\rangle _{\text{%
final}},\left\vert 1\right\rangle _{\text{final}})$ defined by%
\begin{equation}
\left( 
\begin{array}{c}
\left\vert 0\right\rangle _{\text{final}} \\ 
\left\vert 1\right\rangle _{\text{final}}%
\end{array}%
\right) =U\left( 
\begin{array}{c}
\left\vert 0\right\rangle _{\text{initial}} \\ 
\left\vert 1\right\rangle _{\text{initial}}%
\end{array}%
\right) .  \label{Qubit}
\end{equation}%
This unitary matrix defines a one-qubit gate.

\subsection{Phase-shift gate}

We construct the arbitrary phase-shift gate defined by%
\begin{equation}
U_{Z}(\theta )\equiv \text{diag.}(1,e^{i\theta }),  \label{PhaseS}
\end{equation}%
whose action is 
\begin{equation}
U_{Z}(\theta )\left\vert 0\right\rangle =\left\vert 0\right\rangle ,\qquad
U_{Z}(\theta )\left\vert 1\right\rangle =e^{i\theta }\left\vert
1\right\rangle .
\end{equation}%
As is well known, time evolution generates a phase to a state according to
the Schr\"{o}dinger equation. Hence, the two states $|0\rangle $ and $%
|1\rangle $ acquire different phases as time evolves, provided their
energies are made different by the presence of $\mathcal{V}_{1}$. This is
the basic idea of the phase-shift gate.

\begin{figure}[t]
\centerline{\includegraphics[width=0.48\textwidth]{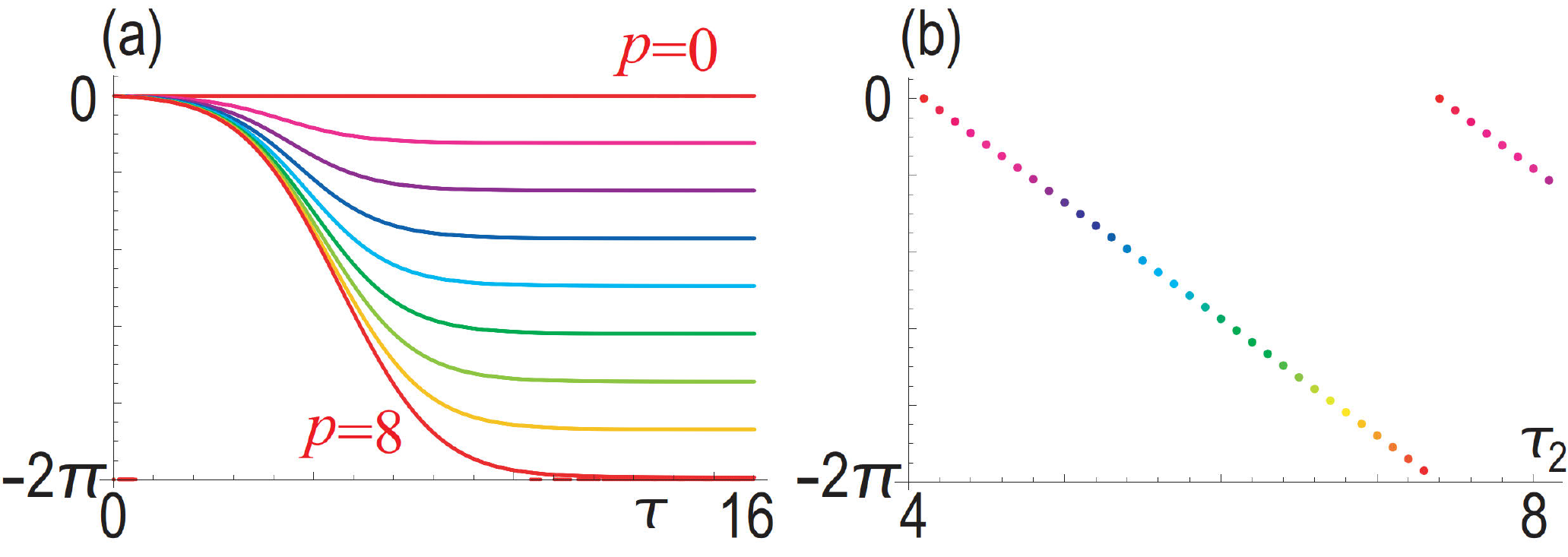}}
\caption{(a) Time evolution of phase modulation. We have set $\protect\tau %
_{1}=4$ and $\protect\tau _{2}-\protect\tau _{1}=0.41p$ with $p=0,1,\cdots
,8 $. (b) Phase modulation as a function of $\protect\tau _{2}$, where $%
4\leq \protect\tau _{2}\leq 8$. We have set $\mathcal{V}_{2}=20$, $\mathcal{V%
}_{1}=1$, $\mathcal{T}=2$ and $T=10$.}
\label{FigPhaseEvolve}
\end{figure}

We temporary control $\mathcal{V}_{1}$ by tuning an applied voltage
difference according to the formula%
\begin{equation}
\mathcal{V}_{1}\left( \tau \right) =\frac{\mathcal{\bar{V}}_{1}}{2}\left[
\tanh \frac{\tau -\tau _{2}}{\mathcal{T}}-\tanh \frac{\tau -\tau _{1}}{%
\mathcal{T}}+2\right] ,
\end{equation}%
with $\tau _{2}\gg \tau _{1}$, while we fix $\mathcal{V}_{2}\neq 0$. We
start either from the state $|0\rangle $ or $|1\rangle $. The absolute value
of the wave function does not change its form but only the phase rotation is
modulated because the state remains in the bottom of the cosine potential.
We show the phase modulation as a function of time for various $\tau _{2}$
in Fig.\ref{FigPhaseEvolve}(a). The phase difference between the initial
state and the final state is shown in Fig.\ref{FigPhaseEvolve}(b), which is
linear as a function of $\tau _{2}-\tau _{1}$. The phase modulations between
the two states $|0\rangle $ and $|1\rangle $ are opposite as in%
\begin{equation}
U_{\theta }\equiv \text{diag.}(e^{-i\theta /2},e^{i\theta /2}),
\end{equation}%
because the energy splitting is opposite between them. Here,%
\begin{equation}
\theta /2\pi =\mathcal{\bar{V}}_{1}f(\mathcal{V}_{2})(\tau _{2}-\tau _{1}).
\end{equation}%
We find $f(\mathcal{V}_{2})=-0.3$\ in the case of $\mathcal{V}_{2}=20$\ by
fitting the line in Fig.\ref{FigPhaseEvolve}(b). This is equivalent to the
phase-shift gate (\ref{PhaseS}), because the overall phase is irrelevant to
quantum gate operations.

\subsection{$\protect\pi /4$ phase-shift gate}

The $\pi /4$ phase-shift gate 
\begin{equation}
U_{T}\equiv \text{diag.}(1,e^{i\pi /4})
\end{equation}
is realized by setting $\theta =\pi /4$ in the generic phase-shift gate (\ref%
{PhaseS}).

\subsection{Pauli-Z gate}

The Pauli-Z gate is realized by the $z$ rotation with the angle $\pi $ as 
\begin{equation}
U_{Z}=-iU_{Z}\left( \pi \right)
\end{equation}%
in the generic phase-shift gate (\ref{PhaseS}).

\subsection{NOT gate}

We construct the NOT gate defined by 
\begin{equation}
U_{\text{NOT}}\equiv \left( 
\begin{array}{cc}
0 & 1 \\ 
1 & 0%
\end{array}%
\right) .  \label{NOT}
\end{equation}%
This gate exchanges the two states $|0\rangle $ and $|1\rangle $. It is
impossible to keep the potential $\mathcal{V}_{2}$ finite since it prohibits
quantum tunneling.

\begin{figure}[t]
\centerline{\includegraphics[width=0.48\textwidth]{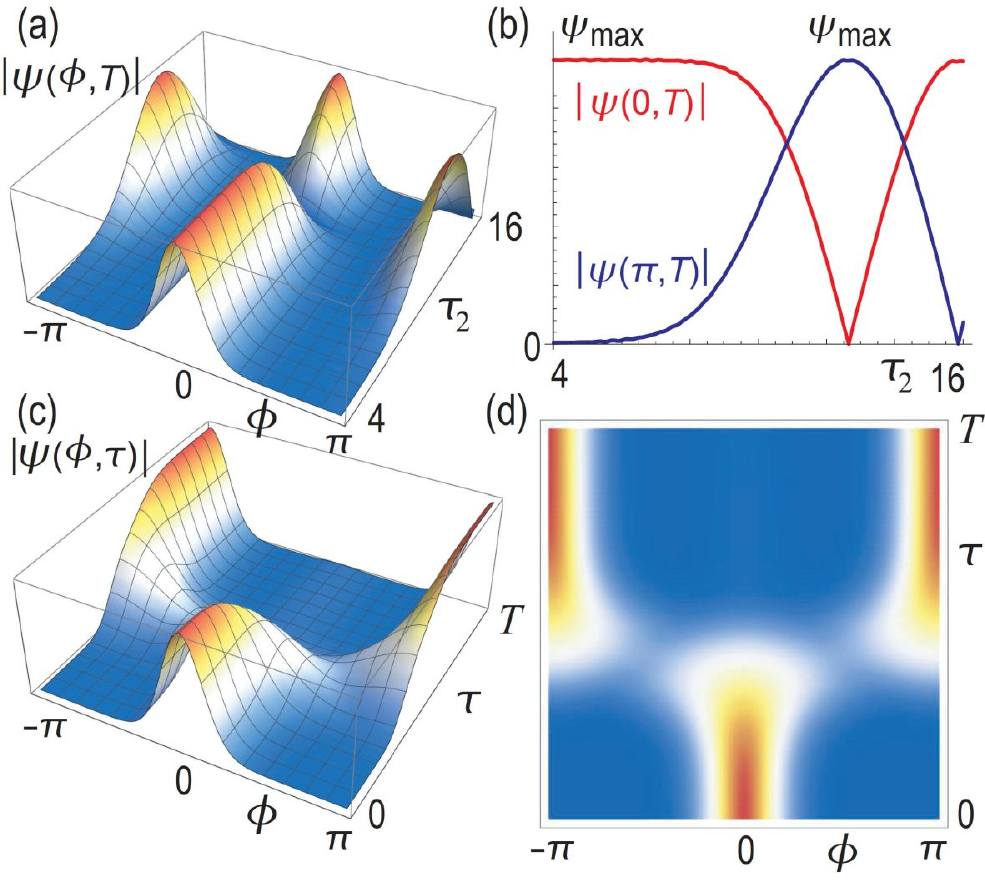}}
\caption{Time evolution of the NOT gate operation. (a) Bird's eye's view of
the final state $|\protect\psi (\protect\phi ,T)|$ as a function of $\protect%
\phi $ and $\protect\tau _{2}$, where $-\protect\pi \leq \protect\phi <%
\protect\pi $ and $4\leq \protect\tau _{2}\leq 16$. (b) The final state $|%
\protect\psi (0,T)|$ colored in red and $|\protect\psi (\protect\pi ,T)|$\
colored in blue as a function of $\protect\tau _{2}$, where $4\leq \protect%
\tau _{2}\leq 16$. (c) Bird's eye's view of $|\protect\psi (\protect\phi ,%
\protect\tau )|$, where $-\protect\pi \leq \protect\phi <\protect\pi $ and $%
0\leq \protect\tau <T$. (d) Top view of $|\protect\psi (\protect\phi ,%
\protect\tau )|$. We have set $\mathcal{\bar{V}}_{2}=20$, $\mathcal{V}_{1}=0$%
, $\protect\tau _{1}=4$, $\mathcal{T}$ $=2$ and $T=20$ in (a) and (b). We
have additionally set $\protect\tau _{2}=12.6$ in (c) and (d).}
\label{FigCosPropagate}
\end{figure}

For this purpose we temporary control $\mathcal{V}_{2}$ by tuning the
applied voltage to the rotator in such a way that%
\begin{equation}
\mathcal{V}_{2}\left( \tau \right) =\frac{\mathcal{\bar{V}}_{2}}{2}\left[
\tanh \frac{\tau -\tau _{2}}{\mathcal{T}}-\tanh \frac{\tau -\tau _{1}}{%
\mathcal{T}}+2\right] ,  \label{PosA}
\end{equation}%
with $\tau _{2}\gg \tau _{1}$, while we set $\mathcal{V}_{1}=0$. The gate
operation requires that the initial state $|0\rangle $ is transferred to the
final state $|1\rangle $ as in Fig.\ref{FigCosPropagate}(c). We explain how
to obtain this time evolution. Let its time evolution be described by the
wave function $\psi (\phi ,\tau )$. The initial condition implies that it
satisfies $\psi (0,\tau )=\psi _{\text{max}}$ and $\psi (\pi ,\tau )=0$ at $%
\tau =0$, where $\psi _{\text{max}}$ is the maximum value of $|\psi (\phi
,\tau )|$. The final state should satisfies $\psi (0,\tau )=0$ and $\psi
(\pi ,\tau )=\psi _{\text{max}}$ at $\tau =T\gg \tau _{2}$, because $%
|0\rangle $ is transformed to the state $|1\rangle $.

This is a nontrivial problem depending on the parameters $\tau _{2}$ in the
applied voltage $\mathcal{V}_{2}\left( \tau \right) $. We fix $\tau _{1}$
arbitrarily and solve the Schr\"{o}dinger equation for $\psi (\phi ,T)$ as a
function of $\tau _{2}$, whose result we show in Fig.\ref{FigCosPropagate}%
(a). There is a certain value of $\tau _{2}$ where $|\psi (0,T)|=0$ and $%
|\psi (\pi ,T)|=\psi _{\text{max}}$, as is clear in Fig.\ref{FigCosPropagate}%
(b). Then, we show the dynamics of $|\psi (\phi ,\tau )|$ with the use of
this value of $\tau _{2}$ in Fig.\ref{FigCosPropagate}(c) and (d), where it
is seen that the initial state $|0\rangle $ localized at $\phi =0$ is
transformed to the final state $|1\rangle $ localized at $\phi =\pm \pi $.
This is the action of the NOT gate.

\subsection{Hadamard gate}

The Hadamard gate is defined by 
\begin{equation}
U_{\text{H}}\equiv \frac{1}{\sqrt{2}}\left( 
\begin{array}{cc}
1 & 1 \\ 
1 & -1%
\end{array}%
\right) .
\end{equation}
It is realized by a sequential application of the Pauli Z gates and the NOT
gate~\cite{Schuch} as%
\begin{equation}
U_{\text{H}}=-iU_{Z}U_{\text{NOT}}U_{Z}.  \label{HZXZ}
\end{equation}

\subsection{Two-qubit gates}

A two-qubit system is made of two rotators put along the $x$ axis as in Fig.%
\ref{FigIllust}. The two-qubit state is expressed as $\left\vert
q_{1}q_{2}\right\rangle $ with $q_{n}=0,1$. An example of the state $%
\left\vert 01\right\rangle $ is given in the system made of Fig.\ref%
{FigIllust}(b3) and (c3). A two-qubit gate operation transforms the initial
state $\left\vert q_{1}q_{2}\right\rangle _{\text{initial}}$ to the final
state $\left\vert q_{1}q_{2}\right\rangle _{\text{final}}$ as%
\begin{equation}
\left( 
\begin{array}{c}
\left\vert 00\right\rangle _{\text{final}} \\ 
\left\vert 01\right\rangle _{\text{final}} \\ 
\left\vert 10\right\rangle _{\text{final}} \\ 
\left\vert 11\right\rangle _{\text{final}}%
\end{array}%
\right) =U\left( 
\begin{array}{c}
\left\vert 00\right\rangle _{\text{initial}} \\ 
\left\vert 01\right\rangle _{\text{initial}} \\ 
\left\vert 10\right\rangle _{\text{initial}} \\ 
\left\vert 11\right\rangle _{\text{initial}}%
\end{array}%
\right) ,
\end{equation}%
which defines the two-qubit gate operation $U$.

\subsection{Two-qubit phase-shift gate}

We apply the voltage difference $V_{12}$\ between the two rotators as in Fig.%
\ref{FigIllust}(b3) and (c3). The potential energy is given by%
\begin{equation}
W\left( \phi _{1},\phi _{2}\right) \equiv W_{2}\left( \phi _{1}\right)
+W_{2}\left( \phi _{2}\right) +\frac{C\left( \phi _{1},\phi _{2}\right) }{2}%
V_{12}^{2},  \label{TetraV}
\end{equation}%
where $W_{2}\left( \phi \right) $ is the potential energy given by Eq.(\ref%
{W2}), and $C(x_{1},x_{2})$ is the capacitance between the rotators,

\begin{equation}
C(\phi _{1},\phi _{2})=\frac{\varepsilon _{0}S}{L\left( \phi _{1},\phi
_{2}\right) }.  \label{Cu}
\end{equation}%
Here, $\varepsilon _{0}$ and $S$\ are the permittivity and the area of the
plates, while $L\left( \phi _{1},\phi _{2}\right) $ is the distance between
the two plates attached to the rotator.

When we apply the Ising gate, the absolute values of the wave functions do
not change but only the phases are modulated. The wave functions are $%
|0\rangle $ or $|1\rangle $, and hence we concentrate on the parallel case $%
\phi _{1},\phi _{2}=0,\pi $. There are relations,%
\begin{align}
L\left( 0,0\right) & =L\left( \pi ,\pi \right) =\ell , \\
L\left( 0,\pi \right) & =2R+\ell ,\quad L\left( \pi ,0\right) =-2R+\ell ,
\end{align}%
where $R$ is the radius of the rotation of the rotator, and $\ell $ is the
length between the supporting points of two adjacent rotators: See Fig.\ref%
{FigIllust}(b3) and (c3).

We calculate%
\begin{align}
W\left( 0,0\right) & =W\left( \pi ,\pi \right) =\frac{\varepsilon _{0}S}{%
\ell }\frac{V_{12}^{2}}{2}\equiv E_{0}, \\
W\left( 0,\pi \right) & =\frac{\varepsilon _{0}S}{\ell +2R}\frac{V_{12}^{2}}{%
2}\equiv E_{+}, \\
W\left( \pi ,0\right) & =\frac{\varepsilon _{0}S}{\ell -2R}\frac{V_{12}^{2}}{%
2}\equiv E_{-}.
\end{align}

We show that the potential energy (\ref{TetraV}) may be written in the form
of the Ising model with field $B_{j}$,%
\begin{equation}
H_{\text{Ising}}=\sum_{j=1}^{N-1}J_{j}s_{j}s_{j+1}+%
\sum_{j=1}^{N}B_{j}s_{j}+E_{0},  \label{IsingA}
\end{equation}%
where $s_{j}=\pm 1$. We rewrite Eq.(\ref{IsingA}) as%
\begin{equation}
H_{\text{Ising}}=\sum_{j=1}^{N-1}H_{j}\left( s_{j},s_{j+1}\right) +\frac{%
B_{1}}{2}s_{1}+\frac{B_{N}}{2}s_{N},  \label{HHs}
\end{equation}%
with%
\begin{equation}
H_{j}\left( s_{j},s_{j+1}\right) =J_{j}s_{j}s_{j+1}+\frac{B_{j}}{2}s_{j}+%
\frac{B_{j+1}}{2}s_{j+1}+\frac{E_{0}}{N-1}.  \label{Hs}
\end{equation}%
We realize the term (\ref{Hs}) by a system made of two adjacent buckled
plates $j$ and $j+1$. There are relations%
\begin{align}
H_{j}\left( 1,1\right) & =W\left( 0,0\right) ,\qquad H_{j}\left(
-1,-1\right) =W\left( \pi ,\pi \right) , \\
H_{j}\left( 1,-1\right) & =W\left( 0,\pi \right) ,\qquad H_{j}\left(
-1,1\right) =W\left( \pi ,0\right) .
\end{align}%
The coefficients in the Ising model are given by%
\begin{align}
J_{j}& =\frac{2E_{0}-E_{+}-E_{-}}{4},  \label{EqB} \\
B_{j}& =-B_{j+1}=\frac{E_{+}-E_{-}}{4},  \label{EqC} \\
E_{0}& =\frac{2E_{0}+E_{+}+E_{-}}{4}.  \label{EqD}
\end{align}

We start with the Gaussian state $\Psi _{\sigma _{1}\sigma _{2}}\left(
x_{1},x_{2}\right) \equiv \psi _{\sigma _{1}}(x_{1})\psi _{\sigma
_{2}}(x_{2})$ with Eq.(\ref{Psi0}) localized at four points $x_{1}=\sigma
_{1}R$ and $x_{2}=\sigma _{2}R$, where $\sigma _{1}=\pm ,\sigma _{2}=\pm $.
The absolute value of this wave function almost remains as it is, but a
phase shift occurs. The unitary evolution is given by 
\begin{equation}
U\left( t\right) =\exp [-i\left( E_{0}/\hbar +\omega \right) t]
\end{equation}%
for $\sigma _{1}=\sigma _{2}=+$\ and $\sigma _{1}=\sigma _{2}=-$, 
\begin{equation}
U\left( t\right) =\exp [-i\left( E_{+}/\hbar +\omega \right) t]
\end{equation}%
for $\sigma _{1}=+$ and $\sigma _{2}=-$, 
\begin{equation}
U\left( t\right) =\exp [-i\left( E_{-}/\hbar +\omega \right) t]
\end{equation}%
for $\sigma _{1}=-$ and $\sigma _{2}=+$, where we have added the zero-point
energy.

It corresponds to the two-qubit phase-shift gate operation,%
\begin{align}
U_{\text{2-phase}}\left( t\right) &=\text{diag.}\left( e^{-i\frac{E_{0}}{%
\hbar }t},e^{-i\frac{E_{-}}{\hbar }t},e^{-i\frac{E_{+}}{\hbar }t},e^{-i\frac{%
E_{0}}{\hbar }t}\right)  \notag \\
&=e^{-i\frac{E_{0}}{\hbar }t}\text{diag.}\left( 1,e^{-i\frac{E_{X}}{\hbar }%
t},e^{i\frac{E_{X}}{\hbar }t},1\right) ,  \label{EqA}
\end{align}%
by identifying the qubit state $\left( \left\vert 00\right\rangle
,\left\vert 01\right\rangle ,\left\vert 10\right\rangle ,\left\vert
11\right\rangle \right) ^{t}=\left( \left\vert ++\right\rangle ,\left\vert
+-\right\rangle ,\left\vert -+\right\rangle ,\left\vert --\right\rangle
\right) ^{t}$.

\subsection{Ising gate}

The Ising gate is defined by $U_{ZZ}\equiv $diag.$(1,-1,-1,1)$, and realized
by setting $E_{X}t/\hbar =\pi $ in Eq.(\ref{EqA}) up to the global phase $%
\exp \left[ -iE_{0}t/\hbar \right] $.

\subsection{CZ gate}

The controlled-Z (CZ) gate is defined by $U_{\text{CZ}}=$diag.$(1,1,1,-1)$.
It is constructed by a sequential application of the Ising gate and the
one-qubit phase-shift gates as\cite{Mak}%
\begin{equation}
U_{\text{CZ}}=e^{i\pi /4}U_{Z}\left( \frac{\pi }{2}\right) U_{Z}\left( \frac{%
\pi }{2}\right) U_{ZZ}.  \label{CZZZ}
\end{equation}

\subsection{CNOT gate}

The CNOT is defined by%
\begin{equation}
U_{\text{CNOT}}^{1\rightarrow 2}\equiv \left( 
\begin{array}{cccc}
1 & 0 & 0 & 0 \\ 
0 & 1 & 0 & 0 \\ 
0 & 0 & 0 & 1 \\ 
0 & 0 & 1 & 0%
\end{array}%
\right) ,
\end{equation}%
and is constructed by a sequential application of the CZ gate (\ref{CZZZ})
and the Hadamard gate $U_{\text{H}}^{\left( 2\right) }$ in Eq.(\ref{HZXZ})
acting on the second qubit as $U_{\text{CNOT}}^{1\rightarrow 2}=U_{\text{H}%
}^{\left( 2\right) }U_{\text{CZ}}U_{\text{H}}^{\left( 2\right) }$.

\subsection{Readout process}

The readout of the plate angle $\phi_n $ can be done for all rotators by
using the fact that the capacitance depends on the relative angle of the two
plates\cite{Bour}. By applying a tiny voltage and by measuring the induced
current, we can readout the capacitance, which is directly related to the
angle $\phi_n $.

\section{Discussion}

\label{SecDiscuss}

We have proposed a universal quantum computer with the use of a chain of
carbon nanotubes together with metal plates attached to them. One-qubit gate
operations are controlled electrically by giving voltage difference between
the attached plate and other plates fixed to the outer system. Two-qubit
operations are controlled electrically by given voltage difference between
the two attached plates belonging to two adjacent nanotubes. We now discuss
the feasibility of such a quantum computer.

We mention experimentally obtained material parameters of a double-wall
nanotube structure\cite{CaiNano}. Nanotube length is 10nm and the intertube
gap length is 0.3nm. The diameter of a nanotube is 10nm\cite{Fenn}. Q factor%
\cite{Papa} is of the order of 100. The inertia $\mu r^{2}$ is 10$^{-30}$kgm$%
^{2}$\cite{Yuz}. The rotational frequency is from 1MHz\cite{Papa} to 100GHz%
\cite{CaiNano}.

If we use a plate with 10nm square with the distance $L=100$nm, the
capacitance $C=\varepsilon _{0}S/L$ is 10$^{-21}$F. If we apply 1mV to the
plate, the electrostatic energy $CV^{2}/2$ is 10$^{-26}$Nm and the operating
time is of the order of 10$\mu $s. If we apply 100mV to the plates, the
electrostatic energy is 10$^{-22}$Nm and the operating time is of the order
of 1ns. These values are experimentally feasible.

This work is supported by CREST, JST (Grants No. JPMJCR20T2).

\end{document}